\documentclass[prd,groupaddress,nofootinbib]{revtex4}

\usepackage{framed}
\usepackage{bbold}
\usepackage{slashed}
\usepackage{graphicx}
\usepackage{amsmath}
\usepackage{amssymb}
\usepackage{epsfig}
\usepackage{hhline}
\usepackage{soul}
\usepackage{tabularx,pifont,multirow}
\usepackage{enumitem}
\usepackage{colordvi}
\usepackage{soul}
\usepackage{xcolor}
\usepackage{yfonts}

\newcommand{\be}{\begin{equation}}
\newcommand{\ee}{\end{equation}}

\definecolor{darkgreen}{rgb}{0,0.3,0.05}
\makeatletter                                                         %
\newcommand*\rel@kern[1]{\kern#1\dimexpr\macc@kerna}                  %
\newcommand*\widebar[1]{                                              %
  \begingroup                                                         %
  \def\mathaccent##1##2{                                              %
    \rel@kern{0.8}                                                    %
    \overline{\rel@kern{-0.8}\macc@nucleus\rel@kern{0.2}}             %
    \rel@kern{-0.2}                                                   %
  }                                                                   %
  \macc@depth\@ne                                                     %
  \let\math@bgroup\@empty \let\math@egroup\macc@set@skewchar          %
  \mathsurround\z@ \frozen@everymath{\mathgroup\macc@group\relax}     %
  \macc@set@skewchar\relax                                            %
  \let\mathaccentV\macc@nested@a                                      %
  \macc@nested@a\relax111{#1}                                         %
  \endgroup                                                           %
}                                                                     %
\makeatother                                                          %

\begin{document}

\preprint[\leftline{KCL-PH-TH/2018-{\bf 62}}

%

\title{\Large {\bf The role of temperature dependent string-inspired CPT violating backgrounds in leptogenesis and the chiral magnetic effect
 } }

\bigskip

\author{Thomas Bossingham}

\affiliation{\vspace{1mm}
Theoretical Particle Physics and Cosmology Group,
  Department of Physics, King's College London, Strand, London WC2R
  2LS, UK}

\author{Nick E.~Mavromatos}

\affiliation{\vspace{1mm} Theoretical Particle Physics and Cosmology Group, Department of Physics, King's College London,  
Strand, London WC2R 2LS, UK}
  
\author{Sarben Sarkar}

\affiliation{\vspace{1mm}
Theoretical Particle Physics and Cosmology Group,
  Department of Physics, King's College London, Strand, London WC2R
  2LS, UK}


\begin{abstract}
\vspace{0.5cm}
\centerline{\bf Abstract }
\noindent\\[-2mm] 

In a \emph{temperature dependent} 
    CPT-Violating (CPTV) axial time-like background (induced by the Kalb-Ramond tensor field of string theory) we discuss leptogenesis by solving the Boltzmann equation.The current work non-trivially modifies the framework of a previous phenomenological approach by the authors where the CPTV \emph{axial} background was considered to be a constant (with no microscopic justification). The constant background approximation though is shown to capture the main phenomenological features of leptogenesis. On comparing our analysis to the related \emph{chiral} magnetic effect for axial current condensates, we conclude that the  Kalb-Ramond field does \emph{not}  play the role of the \emph{chiral} chemical potential needed for that effect. 
 \end{abstract}
\maketitle

\section{Introduction and Motivation \label{sec:intro}}

It has been shown~\cite{ms,decesare,bms} that matter-antimatter asymmetry (through leptogenesis), can occur in appropriate \emph{constant} backgrounds in the cosmological (Robertson-Walker) frame of the early universe.  Such constant backgrounds were associated with postulated axial current condensates.
In the model leptogenesis originates from tree-level decays of a heavy sterile (right-handed, Majorana) neutrino (RHN)  into Standard Model (SM) leptons, in the presence of a generic CPTV time-like axial background \cite{decesare,bms}. The relevant Lagrangian is given by: 
  \be
\label{smelag}
\mathcal{L}= {\mathcal L}_{\rm SM} + i\overline{N}\slashed{\partial}N-\frac{m_N}{2}(\overline{N^{c}}N+\overline{N}N^{c})-\overline{N}\slashed{B}\gamma^{5}N-\sum_k \, y_{k}\overline{L}_{k}\tilde{\varphi}N+h.c.
\ee
where ${\mathcal L}_{\rm SM}$ denotes the SM Lagrangian; $B_\mu$ is a CPTV background field, associated with physics beyond the SM;
$N$ is the RHN spinor field, with (Majorana) mass $m_N$; $N^{c}$ is the charge conjugate spinor;  $\tilde \varphi$ is the adjoint ($\tilde{\varphi}_i=\varepsilon_{ij}\varphi_j $) of the Higgs field  $\varphi$; 
$L_{k}$ is a lepton (doublet) field of the SM sector, with $k$ a generation index; $y_k$ is a Yukawa coupling, which is non-zero and provides a non-trivial (``Higgs portal'') interaction between the RHN and the SM sectors. For simplicity \cite{decesare,bms} we restrict ourselves to the first generation ($k=1$), and set 
\begin{equation}\label{yc1}
y_1 = y ~.
\end{equation}
In \cite{decesare,bms}, the model assumed that $B_\mu$ has only a non-zero temporal component with no time or space dependence (compatible with spatial homogeneity):
\begin{equation}\label{temporalB}
B_0 = {\rm constant} \ne 0~, \, B_i = 0 ~, i=1,2,3~.
\end{equation}
The Lagrangian (\ref{smelag}) then  reduces to a Standard Model Extension (SME) Lagrangian in a Lorentz violating (LV) and CPTV background~\cite{sme}.

In the presence of the background (\ref{temporalB}) and the Higgs portal Yukawa interactions of (\ref{smelag})~\cite{decesare,bms}, a lepton asymmetry is  generated due to the CP and CPTV tree-level decays of the RHN $N$ into SM leptons,:
\begin{eqnarray}\label{4channels}
{\rm Channel ~I}&:& \qquad  N \rightarrow l^{-}h^{+}~, ~ \nu \, h^{0}~,  \\ \nonumber 
{\rm Channel ~II}&:& \qquad  N \rightarrow l^{+}h^{-}~,~  \overline \nu \, h^{0}~.
\end{eqnarray}
where $\ell^\pm$ are charged leptons, $\nu$ ($\overline \nu$) are light ``active'' neutrinos (antineutrinos) in the SM sector,
$h^0$ is the neutral Higgs field, and 
 $h^\pm$ are the charged Higgs fields\footnote{At high temperatures, above the spontaneous electroweak symmetry breaking, the charged Higgs fields $h^\pm$ do not decouple from the physical spectrum, and play an important r\^ole in leptogenesis.}.  As a result of the non-trivial $B_0 \ne 0$ background (\ref{temporalB}), the decay rates of the Majorana RHN between the channels I and II are different, resulting in a lepton asymmetry, $\Delta L^{TOT}$, which then freezes out at a temperature $T_D$. In \cite{bms}, a detailed study of the associated Boltzmann equations for the processes in (\ref{4channels}), and their reciprocals, led to the result:
\be\label{totDL}
\dfrac{\Delta L^{TOT}}{s} \simeq  (0.016, \, 0.019) \,  \dfrac{B_{0}}{m_{N}}, \qquad {\rm at~ freezeout~temperature} \quad T=T_D : \quad m_N/T_D  \simeq (1.44, \, 1.77), 
\ee
where $s$ is the entropy density of the universe.\footnote{} This implies that the phenomenologically acceptable values of the lepton asymmetry of ${\mathcal O}(8 \times 10^{-11})$, can then be communicated to the baryon sector through (B-L) conserving sphaleron processes in the SM (where B is baryon number and L is lepton number).  The observed amount of baryon asymmetry (baryogenesis)  in the universe,  occur for values of
\be\label{b0}
\frac{B_0}{m_N} \sim  10^{-9}, \qquad {\rm at~ freezeout~temperature} \quad T=T_D : \quad m_N/T_D  \simeq (1.77, 1.44),
\ee
With a value of the Yukawa coupling (\ref{yc1}) $y \sim 10^{-5}$, and for $m_N = {\mathcal O}(100)$~TeV~\cite{decesare,bms} we thus obtain a $B_0 \sim 0.1~{\rm MeV}$, for phenomenologically relevant leptogenesis to occur at $T_D \simeq (56 - 69) $ TeV, in our scenario.
In \cite{decesare,bms} the microscopic justification of the background $B_0$ was based on speculation. 
\subsection{Microscopic (string-inspired) framework}

A physically interesting and simple microscopic scenario for $B_0$ is one in which the CPT-Violating background (CPTV) is provided by the field strength of the spin-1 antisymmetric tensor  (Kalb-Ramond (KR)) field which is part of the massless (bosonic) gravitational multiplet of strings~\cite{ms,decesare}. The bosonic gravitational multiplet of a generic string theory consists of three fields~\cite{gsw}: a traceless, symmetric, spin-2  tensor field $g_{\mu\nu}$, that is uniquely identified with the graviton, 
a spin 0 (scalar) field, the dilaton $\Phi$ ( identified with the trace of the graviton), and the spin-1 antisymmetric tensor (Kalb-Ramond) field $B_{\mu\nu} = - B_{\nu\mu}$. In this work we restrict ourselves to the closed string sector, where 
 there is a $U(1)$ gauge symmetry 
$B_{\mu\nu} \rightarrow B_{\mu\nu} + \partial_\mu \theta_\nu - \partial_\nu \theta_\mu$ which characterises the target-space effective action; so in the action it is
the gauge-invariant  three-form field strength of the field $B_{\mu\nu}$,  with components
\begin{equation}\label{hfield}
H_{\mu\nu\rho} = \partial_{[\mu}\, B_{\nu\rho]},
\end{equation}
which appears; the symbol $[\dots ]$ denotes complete antisymmetrisation of the respective indices. 
The 3-form $H_{\mu\nu\rho}$ satisfies, by construction, the Bianchi identity~\footnote{In string theory, in the presence of gauge and gravitational fields, the right-hand-side of (\ref{hfield}) is modified by appropriate Chern--Simons three-forms, which lead to a non-zero right-hand side of the Bianchi identity (\ref{bianchi}), expressing gauge and gravitational anomalies~\cite{gsw}. We shall not deal explicitly with such (higher derivative) terms here, as they are not directly relevant to our leptogenesis scenario.} 
\begin{equation}\label{bianchi}
\partial_{[\mu}\, H_{\nu\rho\sigma]} = 0. 
\end{equation}

The bosonic part of the (3+1)-dimensional effective action, $S_B$, in the Einstein frame is~\cite{string}~\footnote{In this and previous works, our conventions are as follows: space-time metric signature $(+,-,-,-)$, and we shall use the following representation of Dirac $\gamma$-matrices: 
$\gamma^{\mu} =
\begin{pmatrix}
0&\sigma^{\mu}\\
\bar{\sigma}^{\mu}&0
\end{pmatrix}, \;\;
\sigma^{\mu} = 
\begin{pmatrix}
\mathbb{1}\\
\sigma^{\jmath}
\end{pmatrix}, \;\;
\bar{\sigma}^{\mu} = 
\begin{pmatrix}
\mathbb{1}\\
-\sigma^{\jmath}
\end{pmatrix} $, 
$\gamma_{5} = \imath \, \gamma^0\, \gamma^1 \, \gamma^2\, \gamma^3 = \begin{pmatrix}
-\mathbb{1}&0\\
0&\mathbb{1}
\end{pmatrix}.$
}:
\begin{align}\label{sea2}
S_B =&\; \dfrac{1}{2\kappa^{2}}\int d^{4}x\sqrt{-g}\Big(R - e^{-4\Phi}H_{\lambda\mu\nu}H^{\lambda\mu\nu} - \Omega\Big) + \dots,
\end{align}
where  $G= M_P^{-2}$ is the (3+1)-dimensional Newton constant (with $M_P$ the four-dimensional Planck mass), and is related to the string mass scale $M_s$ via~\cite{gsw}: ${G}^{-1} = {\mathcal V}^{(n)} \,  M_s^{2+n}$, with ${\mathcal V}^{(n)}$ a compactification volume (or appropriate bulk volume factor, in brane universe scenarios). For standard (ten space-time dimensional) superstrings n=6. 
The last term $\Omega$ on the rhs of (\ref{sea2}) represents a vacuum energy term. It can arise either in non-critical-dimension string models~\cite{aben}, or from bulk contributions in brane universe scenarios; in the latter case, it includes anti-de-Sitter-type (negative) contributions~\cite{rizos}. The $\dots$ represent terms containing 
derivatives of the dilaton field, $\Phi$; $\Phi$ is assumed~\cite{decesare,bms} to be slowly varying 
at epochs of the Universe, relevant for leptogenesis. As a first approximation we take $\Phi \simeq {\rm constant}$, and absorb it in an appropriate normalisation of the KR field. In this approximation, 
the vacuum energy term $\Omega$ is treated as a constant that is determined phenomenologically by requiring appropriately suppressed vacuum energy contributions.  

It is known~\cite{gsw,string} that the KR field strength terms $H^2$ in (\ref{sea2}) can be absorbed into a generalised curvature scheme with a ``torsionful connection''~\cite{torsion}, with the contorsion proportional to $H_{\mu\nu}^\rho$ field strength, 
${\overline \Gamma}_{\mu\nu}^{\rho} = \Gamma_{\mu\nu}^\rho +  H_{\mu\nu}^\rho  \ne {\overline \Gamma}_{\nu\mu}^{\rho}$,
where $\Gamma_{\mu\nu}^\rho = \Gamma_{\nu\mu}^\rho$ is the torsion-free Christoffel symbol. 
Fermion fields, of mass $m$, are minimally coupled to the contorsion (which is proportional to $H_{\mu\nu}^\rho$ ). The corresponding Dirac term for fermions reads~\cite{kaloper,decesare,bms}:
\begin{align}\label{fermions}
S_{Dirac} &= \, \int d^4x \sqrt{-g} \, \Big[ \frac{\imath}{2} \,\Big(\overline \psi \gamma^\mu {\overline {\mathcal D}}(\overline \omega)_\mu \, \psi - ( {\overline {\mathcal D}}(\overline \omega)_\mu \, \overline \psi  )\, \gamma^\mu \, \psi \Big) - m\, \overline \psi \, \psi \Big], \nonumber \\
& =\; \int d^{4}x\sqrt{-g}\bar{\psi}\Big(\imath\gamma^{\mu}\partial_{\mu} - m\Big)\psi + \int d^{4}x\sqrt{-g} \, ({\mathcal F}_\mu + B_{\mu})\, \bar{\psi}\gamma^{5}\gamma^{\mu}\psi~,  \nonumber \\
 {\overline {\mathcal D}}_a  &= \partial_a  - \frac{\imath}{4} \, \overline \omega_{bca}\, \sigma^{bc}, \quad \sigma^{ab} = \frac{\imath}{2}[\gamma^a, \gamma^b]~,
 \nonumber \\
 {\mathcal F}^\mu & =   \varepsilon^{abc\mu} \, e_{b\lambda} \,  \partial_a \, e^\lambda_c ~,  \quad B^\mu = -\dfrac{1}{4}e^{-2\phi}\varepsilon_{abc}^{\;\;\;\;\;\mu}H^{abc}, \quad J^{5 \mu} = \bar{\psi}\gamma^{\mu}\gamma^{5}\psi,
\end{align}
where $e^{a}_\mu (x)$ are the vielbeins; $g_{\mu\nu} (x) = e_\mu^a (x) \, \eta_{ab} \, e_\nu^b (x)$;  $\eta_{ab}$ is the Minkowski metric of the tangent space at a space-time point with coordinates $x^\mu$;  the generalised spin-connection is: $\overline \omega_{ab\mu}= \omega_{ab\mu} + K_{ab\mu}$;  $K_{abc} =\frac{1}{2} \, (H_{cab}  - H_{abc} - H_{bca}) = - \frac{1}{2} \, H_{abc}$;
$\omega_{ab\mu}$ is the standard torsion-free spin connection\footnote{ The spin connection is given by 
\[{\omega _\mu }^{ab} \equiv e_\nu ^{\;a}\left[ {{\partial _\mu }{e^{\nu b}} + \Gamma _{\;\mu \sigma }^\nu {e^{\sigma b}}} \right],\] with $\Gamma_{\mu\nu}^\lambda = \Gamma_{\nu\mu}^\lambda $ the standard Christoffel symbol.}. Our convention is that Latin letters denote tangent-space indices, while Greek letters refer to space-time indices.  In (\ref{fermions}), we used standard 
properties of the $\gamma$-matrices.  For a Robertson-Walker metric $g_{\mu\nu}$ background, of relevance to us here, ${\mathcal F}_\mu =0$, and thus we can write the 
action (\ref{fermions}) in  the form:
\begin{align}\label{fermions2}
S_{Dirac} = &\; \int d^{4}x\sqrt{-g}\bar{\psi}\Big(\imath\gamma^{\mu}\partial_{\mu} - m\Big)\psi + \int d^{4}x\sqrt{-g}\, B_{\mu} \, \bar{\psi} \gamma^{5}\gamma^{\mu}\psi \, \equiv \; S_{Dirac}^{Free} - \int d^{4}x\sqrt{-g}B_{\mu}J^{5\mu}, 
\end{align}
thus yielding a minimal coupling of the $H_{\mu\nu\rho}$ field to the fermion axial current. 

In four space-time dimensions, the KR three-form $H$ can be expressed in terms of its dual pseudoscalar $b(x)$ (KR ``axion'' ) field~\cite{aben,kaloper}
\begin{align}\label{dual}
\partial^{\mu}b = -\dfrac{1}{4}e^{-2\phi}\varepsilon_{abc}^{\;\;\;\;\;\mu}H^{abc},
\end{align}
where $\varepsilon^{0123} = +1, \; \varepsilon_{0123} = -1$, {\emph etc.} is the gravitationally covariant totally antisymmetric Levi-Civita tensor.
From the definition of $B_\mu$ in (\ref{fermions}), we deduce that 
\be\label{bB}
B^{\mu} = \partial^{\mu}b(x)~.
\ee
The full effective action $S_{eff}$ is given by
\begin{equation}\label{seff}
S_{eff} = S_B + S_{Dirac}.
\end{equation}
A new form of the \emph{effective} action,  in terms of the KR axion field, can be obtained as follows~\cite{kaloper}:
\begin{itemize} 

\item First, we formulate the path integral, which involves a functional integration over the KR field strength $H$ . 

\item We insist on the preservation of the Bianchi identity (\ref{bianchi}) at a \emph{quantum} level, via the addition of appropriate counterterms (in a renormalisation group sense) order by order in perturbation theory. This guarantees the conservation of the ``H-torsion charge ''
$Q = \int d^3 x \, \varepsilon_{ijk} H^{ijk}$, which is implemented in the path-integral by adding a $\delta$-function constraint  in the form $\delta\Big(\kappa^{2}\, \varepsilon^{\mu\nu\rho\sigma} \, \partial_{\mu}\, H_{\nu\rho\sigma}\Big), $  
and expressing the latter in terms of a (dimensionless) Lagrange multiplier field $b(x)$, which eventually will correspond to the dual KR axion field: 
\begin{align}\label{delta}
 \delta ({\kappa ^2}\,{\varepsilon ^{\mu \nu \rho \sigma }}\,{\partial _\mu }\,{H_{\nu \rho \sigma }})
=  \int {\mathcal{D}b} \exp [i\,{\kappa ^{ - 2}}\,\int {{d^4}} x\sqrt { - g} \,b(x){\varepsilon _{\mu \nu \rho \sigma }}{\partial ^\mu }{H^{\nu \rho \sigma }}] \nonumber \\
= \int {\mathcal{D}b} \exp [ - i\,{\kappa ^{ - 2}}\,\int {{d^4}} x\sqrt { - g} \,{\partial ^\mu }b(x){\varepsilon _{\mu \nu \rho \sigma }}\,{H^{\nu \rho \sigma }}]
\end{align}
where the second equality has been obtained by partial integration, upon assuming that the KR field strength dies out at spatial infinity. 

\item Integrating out the $H$-field in the path integral with the action (\ref{seff}), we obtain a path integral over the Lagrange multiplier field $b(x)$, 
\begin{align}\label{seffpi}
Z =&\; \int \, \mathcal{D}g \, \mathcal{D}\psi\, \mathcal{D}\bar{\psi}\, \mathcal{D}b \, \exp[\imath {\tilde S}_{eff}], \nonumber \\
\nonumber
\\
{\tilde S}_{eff} =&\; \dfrac{1}{2\kappa^{2}}\int d^{4}x\sqrt{-g}\,\Big(R + \dfrac{8}{3}\partial_{\sigma} b\, \partial^{\sigma}b - \Omega\Big) 
+ S_{Dirac}^{Free} - \int d^{4}x\sqrt{-g}\partial_{\mu}b\, J^{5\mu}    - \dfrac{3\kappa^{2}}{16}\, \int d^{4}x\sqrt{-g}\,J^{5}_{\mu}J^{5\mu}\Big].
\end{align}

\end{itemize}

In realistic situations there are many fermion species  $\psi_i$,    $i=1, 2, \dots N$ with  masses $m_i$. Then the axial current is a sum over species
\begin{equation}\label{axialcurr}
J_\mu^5 =  \sum_{i}\overline \psi_i \gamma_\mu \, \gamma^5 \, \psi_i ~.
\end{equation}

The reader should notice  the appearance, in the effective action $S_{eff}$ (\ref{seffpi}), of 
a four fermion axial-current-current term, which is a \emph{repulsive} four-fermion term, yielding \emph{de-Sitter type} (positive) contributions to the vacuum energy,
as standard in Einstein-Cartan theories of quantum torsion, where the latter can be integrated exactly in a path integral. \subsection{Choice of background}

Upon splitting the \emph{quantum} field into a background, $\bar b (x)$, and fluctuations, $\tilde b(x)$, 
\begin{equation}\label{split}
b(x) = \bar b (x) + \tilde b(x)
\end{equation}
we then have for $S_{eff}$:
\begin{align}\label{seffbb}
S_{eff} =&\; \dfrac{1}{2\kappa^{2}}\int d^{4}x\sqrt{-g}\,\Big(R + \dfrac{8}{3}\partial_{\sigma} \bar b\, \partial^{\sigma} \bar b - \Omega\Big) 
+ S_{Dirac}^{Free} - \int d^{4}x\sqrt{-g}\partial_{\mu} \bar b\, J^{5\mu}    - \dfrac{3\kappa^{2}}{16}\, \int d^{4}x\sqrt{-g}\,J^{5}_{\mu}J^{5\mu}
\nonumber \\
+& \dfrac{8}{3\,\kappa^{2}}\int d^{4}x\sqrt{-g}\,\partial_{\sigma} \bar b\, \partial^{\sigma} \tilde b
 + \dfrac{1}{2\kappa^{2}}\int d^{4}x\sqrt{-g}\, \dfrac{8}{3}\partial_{\sigma} \tilde b\, \partial^{\sigma}\tilde  b 
+ \int d^{4}x\sqrt{-g}\, \tilde b\, \partial_\mu \, J^{5\mu}  \Big],
\end{align}
where in the last term we performed an integration by parts, for reasons that will become clear below.

We have now reached the point where  crucial assumptions are made which lead to a version of the standard model extension as the starting point of our recent investigations~\cite{ms,decesare,bms} of leptogenesis. 
\begin{itemize}
\item{}

Consider (in the Robertson-Walker frame) KR-axion backgrounds $\bar b(x)$ \emph{linear} in cosmic time $t$, so that ${\dot {\bar b}} $ is 
constant. Such backgrounds rigorously exist in bosonic non-critical strings~\cite{aben} and permit the decoupling of $\bar b$ from $\tilde b$ (since the first term in the second line of (\ref{seffbb}) vanishes as a total derivative
(on assuming that quantum fluctuations $\tilde b$ vanish rapidly at space-time infinity).  
Upon restricting ourselves to the $H$-background terms of the total Lagrangian comprising of the sum of (\ref{smelag}) and (\ref{seffbb}), 
the $\partial \bar b$-$J^5$ interaction term in (\ref{seffbb}) yields the CPT-Violating axial background $B_0$-term of the model discussed in \cite{decesare,bms}, which leads to leptogenesis. In this way one obtains a microscopic origin of $B_0$ in the context of string-inspired models. 

However in our context the existence of such a background remains a postulate. 
 The $b$-axion background linear in time may not constitute exact solutions for superstrings in the presence of fermions. Moreover, even if it was an exact solution,  it is not known whether one could fine tune the associated parameters so as to guarantee a $B_0$ background (\ref{bB}) in the MeV or lower range. We note that in the scenario of \cite{decesare} that a natural mass scale for such backgrounds is provided by the string scale $M_s$ itself and $M_s \gg $~MeV~\cite{aben}. 
\item{}
In \cite{decesare,bms}, another possibility for obtaining a CPTV KR axion background,  corresponding to a constant $B_{0}$ (=${\dot{\bar b}}$),
was proposed. This proposal involves fermionic axial condensates, that have been \emph{conjectured} to occur at the freezeout epoch for the leptogenesis \mbox{scenario} of \cite{decesare,bms}. 
Indeed, in the presence of fermions, the equations of motion for the KR background field $\bar b$ deduced from (\ref{seffpi}), is:
\begin{align}\label{beom}
\partial_{\alpha}\Big[\sqrt{-g}\Big(\dfrac{8}{3\kappa^{2}}\partial^{\alpha}\bar{b} - J^{5 \; \alpha}\Big)\Big] = 0.
\end{align} 
In this proposal we \emph{assume} a (constant) temporal chiral condensate (which respects the spatial isotropy of the universe),
\begin{align}\label{condens}
0 \ne {\rm const}. = \langle J^{05} \rangle = \langle  \psi^\dagger_i \, \gamma^5 \psi_i \rangle.~ 
\end{align}
Such a condensate may characterise fermions in the model except Majorana neutrinos~\cite{decesare}, e.g. quarks in the SM sector; on expanding the current 
in (\ref{beom}) about the condensate (\ref{condens}), $J^{5}_{0}$ = $\langle J^{5}_{0}\rangle + $ quantum fluctuations, and on ignoring the fluctuations, 
we obtain from (\ref{beom})
\begin{align}\label{tempB0}
\partial_{t}\Big[\sqrt{-g}\Big(\dfrac{8}{3\kappa^{2}}B^{0} - \langle J^{0\, 5} \rangle \Big)\Big] = 0,
\end{align}
which allows a solution (\emph{cf.} (\ref{bB}))
\begin{align}\label{b0cond}
B^0 = {\dot {\bar{b}}} = \frac{3\kappa^{2}}{8}\, \langle J_{0\, 5} \rangle = {\rm const.} \ne 0, 
\end{align} 
implying a 
constant  LV and CPTV axial background $B^0$ (in the Robertson-Wallker frame), as required for leptogenesis in the scenario of \cite{decesare,bms}. 

 In the current-era  a plethora of precision measurements~\cite{smebounds}, imply that $|B_0| < 0.01$~eV (as well as  much stronger constraints for the spatial components $|B_i | < 10^{-31}$~GeV). 
In the scenario of \cite{decesare}, this can be guaranteed, if it is assumed that the chiral \mbox{current} condensate $\langle J^{05}\rangle$, is destroyed  at a temperature near the lepton-asymmetry freezeout $T \simeq T_D \simeq 10^5$~GeV  (due to some unspecified physics beyond the SM).
In that case, upon taking into account a Robertson-Walker space-time with scale factor $a(t) \sim T^{-1}$ at high temperatures,  we obtain from (\ref{tempB0}) a cooling `law'  $B_0 \sim T^3$, for $T \lesssim T_D$, which \mbox{comfortably} satisfies the above constraints in the current epoch~\cite{decesare}: the average current temperature of the universe is that of the Cosmic Microwave Background (CMB) radiation, $T_0 \sim T_{\rm CMB} \simeq 0.23$~meV. Indeed, with such a cooling law, taking into account that at \mbox{decoupling} $B_0(T_D \simeq 10^5~{\rm GeV}) ={\mathcal O}(0.1~{\rm MeV})$, one finds~\cite{decesare}: $B_0 (T_0) = {\mathcal O}(10^{-57})$~GeV. 
\end{itemize}
However, the above scenario suffers both from the fact that no concrete model was proposed for the formation of the axial condensates\footnote{Note that the four fermion axial interaction terms in (\ref{seffbb}) are {\emph repulsive}, and as such do not lead to condensate formation. Hence one could invoke other mechanisms, \emph{e.g.} through the appropriate exchange of heavy states that may exist in string theory models. However, such models have not been elaborated further in \cite{bms}.} and the related lack of clarity concerning the nature of the phase transition leading to the disappearance of the condensate soon after the freezeout; at the phase transition the $B_0$ background decreases with the (cubic power of the) temperature~\cite{decesare,bms}, and, as noted above, becomes compatible with the current-era stringent bounds of CPTV~\cite{sme}.  Although the effective theory appears to be a form of Lagrangian in the class of SME, it is \emph{important} to remember that it arises from an underlying ultra-violet complete theory. 

 It is the purpose of this work to offer a \emph{new resolution} to the above issues, by actually considering \emph{non-constant} backgrounds $B_0$, obtained from the antisymmetric tensor field of string theory as described above. This does away with the twin requirements of formation and disappearance of axial current condensates in our earlier works~\cite{decesare,bms}.
  The cubic dependence of temperature for the background $B_{0}\sim T^{3}$ all the way from temperatures around decoupling until the present day, is dictated by the equation of motion of the KR-axion field \emph{in the absence} of an axial-fermion-current condensate; this temperature dependence is sufficiently mild in the high temperature regime of interest, so that the conditions for leptogenesis considered in \cite{bms} are only slightly modified.  
We shall demonstrate that leptogenesis still occurs at decoupling temperatures of order $T_D \simeq 100$ TeV, with the background field, though, smaller than that considered in \cite{decesare}:  $B_0 (T_D) = {\mathcal O}({\rm keV})$; we obtain for the current-epoch value $B_0 (T_0) ={\mathcal O}(10^{-59})$~GeV, which lies comfortably within the stringent current bounds of CPTV and LV~\cite{smebounds}.   
  
Before proceeding, we would like to mention another important issue. Since our effective field theory appears to be one with a chiral chemical potential provided by the KR torsion field  $B_0$, on minimally extending the model by adding an external magnetic field, one would be tempted to conjecture that  the conditions for the chiral magnetic effect (CME)~\cite{cme} would be satisfied; however, this is not so. The $B_0$ field is actually a fully-fledged  axial background rather than a mere chiral chemical potential, and it is known that such backgrounds make no contributions to the CME~\cite{kaplan,dvorn}~\footnote{Explicitly, constant backgrounds, unlike the chemical potentials, change the fermion dispersion relation. This is a crucial difference.}. The CME is connected with quantum (chiral~\cite{adler}) anomalies, and, as we mentioned previously, in our string-inspired effective theory, the field $B_0$ is associated with the KR H-torsion. It is well known~\cite{hull,mavindex}, that the contributions of the latter can be removed from the anomaly equation by an appropriate choice of a renormalisation-group scheme. Hence, there should be 
no physical effects of the KR torsion field $B_0$ on the anomaly equation, and thus on the CME~\footnote{If the CME had been present then this would have opened up the possibility of a new mechanism for leptogenesis due to primordial magnetic fields.}.  For completeness we shall give a fuller discussion of CME in the appendix.

\section{Temperature-dependent background field $\mathbf{B_0(T)}$ \label{sec:bocond}}

Without assuming the formation of constant chiral condensates let us consider the \mbox{equation} of motion (\ref{tempB0}) for the KR-axion background field $\bar b (x)$ (\ref{bB}). We next replace the condensate of the axial current  in (\ref{tempB0}) by its thermal counterpart, $\langle J^{0\, 5}\rangle_T$, since we assume thermal equilibrium for (high) temperatures above decoupling of the heavy sterile neutrinos $T \geq T_D$,. Contrary to the scenario of \cite{bms}, however, we shall assume a $T$-dependent $B_0(T)$. In our cosmological scenario, the backgrounds depend at most on cosmic time, which in turn can be related to temperature, $t$ = $t(T)$. 
The relationship between the cosmic time $t$ and temperature $T$   \mbox{depends} on the cosmological era. If we assume that the decoupling temperature $T_D ={\mathcal O}(100)$~TeV (as was the case for  the constant-$B_0$ case of  \cite{bms})  the relevant cosmological era is the radiation era. This assumption will, a posteriori, be shown to be consistent. In the radiation era, 
for which the scale factor $a(t)$ of the universe scales as follows :
\begin{equation}\label{tT}
a(t)_{\rm rad} \sim t^{1/2} \sim T^{-1} \, \Rightarrow \, t \sim T^{-2}. 
\end{equation}
The metric determinant then scales in that era as $\sqrt{-g} \propto a(t)^3 \propto T^{-3}$. 

The most general  solution  of (\ref{tempB0}) then, when expressed in terms of $T$ reads:  
\begin{align}\label{tempB02}
\dfrac{8}{3\kappa^{2}}\, B^{0}(T) - \langle J^{0\, 5} \rangle_T  = \frac{{\rm constant}}{\sqrt{-g(t(T))}} = A^\prime\, T^3,
\end{align}
where $A^\prime$ is a constant of integration,  and we took into account that both $B_0$ and $\langle J^{0\, 5}\rangle_T$ are only functions of time (or equivalently, temperature).

Our next step is to evaluate the thermal condensate of the axial current, entering (\ref{tempB02}), using 
equilibrium thermodynamics in the presence of the background $B_0(T)$. 
Upon expanding the chiral current around the thermal average (or condensate) plus terms dependent on fermion excitation fields, $\psi_i$, we have
\begin{align}\label{expnsg5}
J^{0\, 5} = \langle J^{05} \rangle_T + \sum_{i={\rm fermions}}\,  \psi^\dagger_i \, \gamma^5 \, \psi_i .
\end{align}
We shall later show that all the contributions to the thermal current condensate exactly cancel. 
 The fermion part of the Lagrangian  
reads:
\begin{align}\label{fermlag}
\mathcal{L}^F =\; \sqrt{-g}\bar{\psi}\Big[\imath\gamma^{\mu}\partial_{\mu} - m\mathbb{1} - (B_{0}(T) - \mu_{5}(T))\gamma^{0}\gamma^{5} + \mu (T)\gamma^{0}\Big]\psi
+ \dots 
\end{align}
with 
\begin{align}\label{m5}
\mu_5(T) =   \dfrac{3\kappa^{2}}{8}\,  \langle J^{05}\rangle_T~,
\end{align} 
and the $\dots$ denoting four fermion interactions (\emph{cf.} (\ref{seffbb})).  
For reasons that will be clarified in the appendix, we have also added a chemical potential $\mu (T)$ for the fermions. 
In phenomenologically realistic situations,  $\mu$ is taken to be non-zero \emph{only} for the quark fields of the SM; since lepton number is not conserved, the lepton chemical potentials are zero.  

As we shall demonstrate below, in the absence of external electromagnetic fields,  
\begin{align}\label{m5vanish}\langle J^{05}\rangle_T = 0~.
\end{align}
To evaluate $\langle J_0^5 \rangle_T$   we concentrate on the free part of the Lagrangian (\ref{fermlag}), ignoring the four-fermion axial-current-current interactions appearing in (\ref{seffbb}); the current-current interactions are suppressed by the gravitational coupling $\kappa^2  \propto M_P^{-2}$ (with $M_P$ the Planck mass). This suppression is essential in
obtaining the dispersion relation for the fermion fields -it will be convenient to consider a generic fermion of mass $m$ with helicity $\lambda$ (and chemical potential $\mu$ in the thermal distribution):
\begin{align}\label{dr}
 E = \sqrt{[p + \lambda\, B_{0}]^{2} + m^{2}}. 
\end{align}

Taking the massless limit, $ m \to 0$, appropriate for the high temperature regime $ T \gtrsim T_D ={\mathcal O}(100)~{\rm TeV}$, we have
\begin{align}\label{disp}
E = | p + \lambda B_0 |
\end{align}
For anti-particles the corresponding dispersion relation is obtained from (\ref{disp}) by the replacement 
$E $$\rightarrow$$ \,- E$. The antiparticles are also characterised by the substitution \mbox{$\mu, \mu_5 \to - \mu, -\mu_5$}. 

 The thermally-averaged axial current $\langle J_0^5 \rangle_T$ is the sum over all fermion species (leptons and quarks) (\emph{cf.} (\ref{axialcurr})), hence we obtain for its thermal average
\begin{align}\label{tac}
\langle J^{5}_{0}\rangle_T &=\sum_{i} \langle\psi^\dagger\, \gamma^{5}\, \psi_{i}\rangle_T  = -\sum_{i,\lambda}\Big[(n^{eq}_{L} - \bar{n}^{eq}_{L}) - 
(n^{eq}_{R} - \bar{n}^{eq}_{R})\Big]_{i,\lambda} \nonumber \\ &= -\sum_{i,\lambda} \, \int \frac{d^3p}{(2\pi)^3} \, \Big[ \Big(f^{eq}(E_L, \mu_L) 
- f^{eq}(E_L, -\mu_L) \Big) - (L \leftrightarrow R) \Big],
\end{align} 
where $n^{eq}$ ($\bar{n}^{eq}$) is the (thermal equilibrium) number density of particles (anti-particles), and the sum is still over all fermion species $i$, and over  both helicities $\lambda =\pm 1$. The (left-(L), right-(R)handed) thermal distributions are $f(E_{L,R}, \mu_{L,R})$, with $f(E, \mu)=\frac{1}{e^\frac{E-\mu}{T} + 1}$,  and $\mu_{L(R)}= \mu - (+)\mu_5$. It is straightforward to see from (\ref{tac}), then, that there is an exact cancellation among the various terms appearing on the right-hand side, since in the massless chiral fermion case the right-handed and left-handed particles swap over when considering the opposite helicity.  Hence, the thermal expectation value of the axial current (\ref{tac}) vanishes 
(\ref{m5vanish}). In the presence of external electromagnetic fields, though, this is \emph{not} the case, due to (quantum) chiral anomalies, as we shall discuss later on in the appendix.

Therefore the thermal current expectation value does not contribute to $B_0(T)$ and the temperature dependence of the background field behaves like, 
\begin{align}\label{B0sol}
 B_{0}(T) \simeq  A \, T^3, \quad A \equiv \frac{3\kappa^2}{8}\, A^\prime,
\end{align}
where the constant $A$ will be determined by the boundary condition  $B_{0}(T = T_{D}$$ \sim100~{\rm TeV})$, which in turn is given by the requirement of the production of phenomenologically acceptable 
values for lepton asymmetry. 

Before proceeding to leptogenesis, we comment at this stage on the approximate constancy of the background $\bar b$ in the lepton-asymmetry decoupling era, when $T \simeq T_D \sim 10^5$~GeV, as implied by the scaling (\ref{B0sol}). From (\ref{tT}), we can easily evaluate its rate of change with the cosmic time as measured in (natural) units of $M_P^{-1}$:
\begin{align}\label{bdotb}
M_P^{-1} \, \frac{{\dot {\bar b}}}{b} = -3\, \frac{H(t)}{M_P} \, \ll 1 ,  \quad T(t) \simeq T_D
\end{align}
where $H(t)$ is the Hubble rate, and we took into account that this is much smaller than $M_P$ during the leptogenesis era\footnote{In models based on string theory/Grand Unification (GUT) we note that $H = {\mathcal O}(10^{-4}-10^{-5})\, M_P$ during the GUT era, $T\sim 10^{14}-10^{16}$~GeV, after the inflation exit.}. Thus, our assumption of treating $B_0$ as approximately constant when evaluating the lepton asymmetry (as we shall do in the following section), appears to be self consistent. 

\section{Leptogenesis \label{sec:boltz}} 


In this section we proceed to solving the system of the pertinent Boltzmann equations that will allow a determination of the abundance of the heavy neutrino and the lepton asymmetry~\cite{bms}.  The two Boltzmann equations, of relevance to us here, are given below for the decay processes into the charged leptons $N \leftrightarrow l^{-}h^{+}$ and $N \leftrightarrow l^{+}h^{+}$, as well as the decay processes into the neutral leptons $N \leftrightarrow \nu h^{0}$ and $N \leftrightarrow \bar{\nu}h^{0}$. The heavy neutrino Boltzmann equation is given by,
\begin{align}
\dfrac{d\bar{Y}_{N}}{dx}&\; + P(x)\bar{Y}_{N} = Q(x), \;\;\;\;\;\;\;\; x \equiv \frac{m_N}{T}  < 1\\
\nonumber
\\
\nonumber
P(x) =&\; 2a^{2}x^{10/3}(1 - 0.3909x^{2} + 0.2758x^{4}), \;\;\;\;\;\;\;\; a^{2} \equiv \dfrac{0.0724e\vert y\vert^{2}M_P}{168g_{N}\pi m_{N}} \simeq 2.2763\\
\nonumber
Q(x) =&\; 2b^{2}x^{10/3}(1 - 0.5668x^{2} + 0.3749x^{4}), \;\;\;\;\;\;\;\; b^{2} \equiv \dfrac{0.0957\vert y\vert^{2}M_P}{168(2\pi)^{3}m_{N}} \simeq 0.028.
\end{align} 
The lepton asymmetry Boltzmann equation is given by,
\begin{align}
\dfrac{d\mathcal{L}}{dx} +&\; J(x)\mathcal{L} = K(x), \;\;\;\;\;\;\;\; x < 1\\
\nonumber
\\
\nonumber
J(x) =&\; \omega^{2}x^{10/3}(1 - 0.5668x^{2} + 0.3749x^{4}), \;\;\;\; \omega^{2} \equiv \dfrac{0.0362e\vert y\vert^{2}M_P}{84g_{l}\pi m_{N}} \simeq 1.1381,\\
\nonumber
\\
\nonumber
K(x) =&\; \Big[\nu^{2}x^{13/3}(1 - 0.2385x^{2} - 0.3538x^{4})\bar{Y}_{N}(x) - \sigma^{2}x^{13/3}(1 - 0.1277x^{2} - 1.4067x^{4}) - \delta^{2}\Big]\dfrac{B_{0}(x)}{m_{N}}\\
\nonumber
\\
\nonumber
\nu^{2} \equiv&\; \dfrac{0.1041e\vert y\vert^{2}M_P}{84g_{N}\pi m_{N}} \simeq 6.5459, \;\;\;\; \sigma^{2} \equiv \dfrac{0.0479\vert y\vert^{2}M_P}{84(2\pi)^{3}m_{N}} \simeq 0.0281, \;\;\;\; \delta^{2} \equiv \dfrac{21.4104}{84}\dfrac{g_{l}}{\pi^{2}e} \simeq 0.038.
\end{align}
where for $0 < x < 1$ the solution for $B_0(x) $ is given by (\ref{B0sol}):
\begin{align}\label{B0solz}
B_0 = A\, T^3 = \Phi \, x^{-3}, \quad   \Phi  \equiv A \, m_N^3. 
\end{align}
This is the modification from the analysis given in \cite{bms}. The integrating factor method (for a linear first order differential equation) is used to solve the set of differential equations. For the heavy neutrino the general solution is given by,
\begin{align}
\bar{Y}_{N}(x) =&\; I_{N}^{-1}(x)\int^{x}d\tilde{x}I_{N}(\tilde{x})Q(\tilde{x})
\end{align}
the integrating factor is calculated to be,
\begin{align}
I_{N}(x) =&\; D\exp\Big[x^{13/3}(1.0506 - 0.281x^{2} + 0.1507x^{4})\Big].
\end{align}
where during the proceeding calculating the exponential from the integrating factor is expanded to first order. The general solution for the heavy neutrino Boltzmann equation is given by,
\begin{align}
\nonumber
\bar{Y}_{N}(x < 1) \simeq&\; \Big[1 - x^{13/3}(1.0506 - 0.281x^{2} + 0.1507x^{4})\Big]\\
\nonumber
\times&\; \Big[x^{13/3}(0.0129 - 0.005x^{2} + 0.0025x^{4} + 0.0068x^{13/3} - 0.0046x^{19/3}\\
\nonumber
+&\; 0.0031x^{25/3} - 0.0007x^{31/3} + 0.0002x^{37/3}) + 0.0123\Big],
\end{align}  
where the constant of integration was found by taking the limit $x \rightarrow 0$ and equating the above expression to the equilibrium abundance $\bar{Y}_{N}(x \rightarrow 0) \rightarrow 0.0123$. The lepton asymmetry Boltzmann equation is solved in a very similar way; the general solution is given by
\begin{align}
\mathcal{L}(x) =&\; I_{\mathcal{L}}^{-1}(x)\int^{x}d\tilde{x}K(\tilde{x})I_{\mathcal{L}}(\tilde{x}).
\end{align}
where the integrating factor is given by,
\begin{align}
I_{\mathcal{L}}(x) =&\; \exp\Big[\int^{x}d\tilde{x}J(\tilde{x})\Big] = \tilde{D}\exp\Big[0.2626x^{13/3} - 0.1019x^{19/3} + 0.0512x^{25/3}\Big].
\end{align}
Again when using the integrating factor, the expression is simplified by expanding the exponential to first order in the calculation of the integral appearing in the general solution. The general solution is given by, 
\begin{align}
\mathcal{L}(x < 1) \simeq&\; \Big[1 - x^{13/3}(0.2626 - 0.1019x^{2} + 0.0512x^{4})\Big]\\
\nonumber
\times &\;\Big[0.019x^{-2} + 0.0182x^{7/3} - 0.0027x^{13/3} + 0.0014x^{19/3} + 0.002x^{20/3} - 0.0023x^{26/3} + 0.0013x^{32/3}\\
\nonumber
-&\; 0.004x^{11} + 0.0001x^{38/3} + 0.0028x^{13} - 0.0001x^{44/3} - 0.0005x^{15} - 0.0038x^{46/3} + C_{0}\Big]\dfrac{\Phi}{m_{N}}
\end{align}
the expression above has been truncated to the same order as the solution for the heavy neutrino abundance. The final constant of integration ($C_{0}$) is found by equating the lepton asymmetry equilibrium abundance to the Boltzmann solution ($\mathcal{L}^{eq}(x_{P}) \simeq \mathcal{L}(x_{P})$) evaluated at a point $x_{P} < 1$.
\begin{align}
\mathcal{L}^{eq}(x) \simeq 0.0455\dfrac{B_{0}(x)}{m_{N}}x, \;\;\;\;\;\;\;\;\;\;\;\; B_{0}(x) \simeq \Phi x^{-3},
\end{align}
where the lepton symmetry equilibrium abundance is the difference between the leptons of helicity $\lambda = -1$ and the anti-leptons of helicity $\lambda = +1$. Once a point $x_{P}$ is chosen we perform a $[7,  7]$ diagonal Pad\'e expansion around this point for the two Boltzmann equation solutions ($\mathcal{L}(x < 1)$ and $\bar{Y}_{N}(x < 1)$) in order to study the regime of $x = x_{D} \geq 1$. The observable lepton asymmetry is then calculated by,
\begin{align}
\dfrac{\Delta L^{Total}}{s} = \dfrac{\mathcal{L}^{P}(x_{D} = 1, \; x_{P} < 1)}{2\bar{Y}_{N}^{P}(x_{D} = 1, \; x_{P} < 1)} = q\dfrac{\Phi}{m_{N}} \simeq 8 \times 10^{-11},
\end{align}
where $q$ is a number that differs depending on the expansion point $x_{P}$. Equating this result to the phenomenological value we obtain a range of values for the constant $\Phi$ depending on the chosen point $x_{P}$, the value of decoupling is $x_{D} = 1$. The results are given in table \ref{table1}. 
 
\begin{table}[ht]
\caption{A table showing the different values of the constant $\Phi$ and the background field $B_{0}$ with respect to different expansion points $x_{P}$ at the decoupling value of $x_{D} = 1$.}
\begin{tabular}{p{3cm}p{3cm}p{3cm}p{3cm}}
$x_{P}$ &  $\dfrac{\Phi}{m_{N}}$  &  $\Phi$(keV)  &  $B_{0}(x_{D})$(keV)\\
&&&\\
\hline\hline
&&&\\
 0.50 & $3.6\times 10^{-12}$ & 0.36 & 0.36 \\
 0.75 & $5.9\times 10^{-12}$ & 0.59 & 0.59 \\ 
 0.90 & $7.4\times 10^{-12}$ & 0.74 &  0.74
\end{tabular}
\label{table1}
\end{table}
  
Compared to the case of constant $B_0$ studied in \cite{bms}, we observe that the LV and CPTV value of the background field $B_0$ at decoupling 
$T_D=\mathcal {O}(100)$ TeV, which yields phenomenologically acceptable lepton asymmetry in the universe is smaller, is in the keV range. 
  
\section{Current-era magnitude of CPTV Background and Vacuum (Dark) Energy Contributions \label{sec4}}

In this section, the magnitude of the background field is studied for temperatures below the point of decoupling $T < T_{D}$ (which correspond to $x > 1$). The quarks fall out of equilibrium at $x \simeq 578$. For larger $x$ values the current expectation value vanishes. The $x$-value today is $x = x_0 \simeq 4.2\times 10^{17}$. The behaviour of the background field around the point of decoupling and beyond is given by,
\begin{align}
B_{0}(x) \simeq \Phi x^{-3}, \;\;\;\;\;\;\;\;\;\;\;\; \Phi \simeq (0.36 - 0.74) \; {\rm keV},
\end{align}
where the constant $\Phi$ was found using the leptogenesis calculation. The results are shown in  table \ref{table2}.

\begin{table}[ht]
\caption{A table showing the different values of the background field $B_{0}$ in the temperature regimes where the quarks fall out of equilibrium (first line) and the value today (second line).}
\begin{tabular}{p{3cm}p{3cm}p{3cm}p{3cm}}
$T$(eV) &  $x$  &  $B_{0}$(eV)\\
&&\\
\hline\hline
&&\\
 $173\times 10^{9}$ & 578 & $(1.9 - 3.8)\times 10^{-6}$ \\
 $2.4\times 10^{-4}$ & $4.2\times 10^{17}$ & $(4.9 - 10.0)\times 10^{-51}$
\end{tabular}
\label{table2}
\end{table}

These values indicate that the current value of the ``torsion'' LV and CPTV field $B_0$ lies comfortably within the current bounds~\cite{smebounds}, $B_0< 0.01$~eV and (for the spatial components) $B_i < 10^{-31}$~GeV; so even a boost by small velocities, due to a difference of the laboratory frame with respect to the cosmological frame, will still yield spatial components within the above limits.

However, in contrast to the standard SME case, where $B_0$ appears only as a background field coupled to axial fermion currents, in our microscopic formulation there is also a cosmological vacuum energy density contribution $\rho^{{\rm D.E.}}_{B_0}$ due to the kinetic term of the ${\dot {\bar b}} \equiv B_0$ field in the Lagrangian (\ref{seffbb}):
\begin{align}
\rho^{{\rm D.E.}}_{B_0} (x) = \dfrac{4}{3\kappa^{2}}\partial_{\mu}\bar{b}\, \partial^{\mu}\bar{b} = \dfrac{M_P^{2}}{6\pi}B_{0}^{2}(x) = \dfrac{M_P^{2}\Phi^{2}}{6\pi}x^{-6} 
\end{align}
As we see from Table (\ref{table2}), such dark energy contributions (scaling with the temperature as $T^6$) assume the value $\rho^{{\rm D.E.}}_{B_0} (x_D=1) \simeq (0.5 - 1.9) \times 10^{-52} \; M_P^{4}$ at decoupling $x_D =1$ (\emph{i.e.} $T_D ={\mathcal O}(10^5)$\,GeV), while today 
yield a vacuum energy density of order $ \rho^{{\rm D.E.}}_{B_0} (x_0) \simeq 10^{-158}~M_P^4$ today, which is well within the cosmologically observed current value  $10^{-122}~M_P^4$~\cite{planck}. 

\section{Conclusions} 

In this work we have generalised our previous study of leptogenesis based on a  \emph{constant} LV and CPTV time-like axial background  to the case of a torsion background  varying with the temperature of the early universe.  The torsion is provided here by the antisymmetric tensor (Kalb-Ramond) spin-one field of the massless bosonic multiplet of closed string theory.  The \emph{phenomenology} of our leptogenesis, remains largely unchanged from the constant background case, and is consistent with the stringent current epoch constraints on LV and CPTV, as well as cosmological constraints on the vacuum energy density.

Before closing, we would like to remark that during the leptogenesis era, there might be present primordial external magnetic fields, which can also lead to 
leptogenesis, however  via mechanisms which are distinct from the one in our work~\cite{maglept}. In the presence of a chiral chemical potential $\mu_5$, that is a difference of the chemical potentials $\mu_L - \mu_R$ between left(L)- and right(R)-handed spinors, it is known that one has an induced electric current proportional to the magnetic field intensity, the phenomenon of CME~\cite{cme}. In addition to 
primordial eras, of relevance to leptogenesis, systems such as neutron stars or a hot QCD quark-gluon (QG) plasma with external magnetic fields  show the CME.  Given that our temperature dependent axial background (torsion) field $B_0(T)$ survives until today, and its coupling to a chiral fermion current in the effective action has the apparent form of a chiral chemical potential term, albeit temperature dependent, it is natural to examine whether the axial background $B_0$ has any effect on the CME. 

This question is examined in the appendix. As we shall see, though, in our case, the CME is not generated by the presence of the axial KR background $B_0$ (which in this respect plays a r\^ole analogous to an external axial vector potential that is known not to contribute to CME~\cite{kaplan,dvorn}). The non-contribution of the $B_0$ field to the CME in our case should also be expected from:
\begin{enumerate}
  \item {the fact that the phenomenon
 has its origin~\cite{cme} in the chiral anomalies of quantum field theory~\cite{adler}}
  \item{the r\^ole 
of the $B_0$ KR field as a  torsion in the low-energy string effective action}
  \item{the well-known result~\cite{hull,mavindex} that  torsion contributions to the anomaly equation are removable by the addition of appropriate local counterterms (in a renormalisation group sense) to the corresponding effective action. Physical effects, such as the CME, should thus be free from such ambiguities.} 
\end{enumerate} 

This result invalidates any arguments~\cite{dvorn2} in favour of the axial background $B_0$ playjng a r\^ole in the generation of instabilities and thus magnification of the magnetic fields in neutron stars, though it must be said that the KR torsion might play a non-trivial r\^ole in the dynamo equation for the generation of magnetic fields~\cite{dynamo}, and thus affect their strength in a way independent of the CME~\footnote{We remark at this point that 
the claimed CME-like effect in ref.~\cite{dynamo}, due to the axial component of the (generic) torsion, $T_\mu = \epsilon_{\mu\nu\rho\sigma}K^{\nu
\rho\sigma}$, with $K^{\nu\rho\sigma}$ the contorsion tensor, arises because of an extension of the torsionful model considered in that work, which entails the 
replacement of the partial derivative $\partial_\mu$ in the divergence of the axial current entering the anomaly equation by the 
quantity  $D_\mu = \partial_\mu - \imath \, T_\mu$. However, 
such a prescription is not available within the framework of our current work, where the anomaly equation has the form (\ref{anomalygrav}) (see appendix). The gravitational covariant derivative entering the divergence of the axial four-current in the anomaly equation in a curved space-time is necessarily torsion free, for symmetry reasons, as we will discuss in the appendix.}. We hope to come back to a discussion of such effects in a future work.

\appendix

\section{Torsion, Anomalies and Chiral Magnetic Effect \label{sec:cme}} 

\numberwithin{equation}{section}

\setcounter{equation}{0}

In this appendix we are interested in examining whether our UV complete microscopic model which has a $B_0$ that formally plays the r\^ole of a chiral chemical potential $\mu_5$ displays a chiral magnetic effect~\cite{cme}.
The answer to this question is negative, as expected for reasons stated in the concluding section of the article. Here we give some details towards a mathematical proof within the framework of the low-energy effective quantum field theories obtained from strings. 

 It will be instructive to first review briefly the CME phenomenon in the QG plasma case~\cite{cme}. Consider the (3+1)-dimensional flat-space-time, finite-density \emph{massless} quark Lagrangian in the presence of a finite chemical potential $\mu$ and a finite chiral chemical potential $\mu_5$ which we will denote by 
\begin{align}\label{chirallag}
{\mathcal L}_{\rm quarks} \,  \ni \, \int \, d^4 x \, \Big(\mu \sum_{i={\rm quarks}} \, {q}_i^\dagger \, q_i  + \mu_5 \sum_{i={\rm quarks}} \, 
{q}_i^\dagger \, \gamma^5 \, q_i  \Big).
\end{align}
The chiral anomaly implies that the the corresponding chiral current density $J^{5\, \mu}$ is \emph{not conserved}, but its divergence is given by the so-called axial~anomaly~\cite{adler}, which in the case of interest is restricted only to include electromagnetic terms with Maxwell field strength $F_{\mu\nu}= \partial_\mu\, A_\nu - \partial_\nu \, A_\mu$ (with  $A_\mu$ denoting the U(1) gauge potential, corresponding to the photon field), and its dual $\widetilde F_{\mu\nu} = \frac{1}{2}\, \epsilon_{\mu\nu\alpha\beta}\, F^{\alpha\beta}$ (with $\epsilon^{0123}=+1$ in our conventions): 
\begin{align}\label{anomaly}
\partial_\mu J^{5\, \mu} = \frac{e^2}{8\, \pi^2} \, F_{\mu\nu} \, \widetilde F^{\mu\nu} = \frac{e^2}{2\, \pi^2} \, \vec E \cdot \vec {\mathcal B}~,
\end{align}
where $e$ is the electron charge, and $\vec E$ ($\vec {\mathcal B}$) is the electric (magnetic) field respectively, which will be taken to be external in our  discussion. 

Integrating over 3-space, we may rewrite (\ref{anomaly}) in terms of the rate of change of the \emph{chirality} $N_5 = N_R - N_L$~\cite{cme}:
\begin{align}\label{chiral2}
\mu_5 \, \frac{d N_5}{dt} = \frac{e^2\, \mu_5}{2\, \pi^2} \, \int d^3x \, \vec E \cdot \vec {\mathcal B}~.
\end{align}
In arriving at (\ref{chiral2}) we took into account  that the chiral chemical potential $\mu_5$ is the energy required to change a left handed fermion into a right handed one, which equivalently is the energy required to move a particle from the left handed Fermi surface and place it onto the right-handed one. If $\mu_{L(R)} = \mu \mp \mu_5$ denotes the corresponding chemical potentials of the left(right) handed fermions, the above process  costs an energy~\cite{cme} $\mu_R - \mu_L = 2\mu_5$, and this will change the chirality $N_5$ by 2. For an infinitesimal change $dN_5$ of the chirality then, the corresponding cost in energy is given by $\mu_5\, dN_5$, whose rate is then given by (\ref{chiral2}). 
Conservation of energy, implies that this amount must be compensated by the power of the electric field present in the system, which in terms of the electric current density $\vec j_E$  is provided by $\int d^3x \, \vec j_E \cdot \vec E$, thereby leading (on account of (\ref{chiral2})) to:
\begin{align}\label{chiral3a}
\int d^3x \, \vec j_E \cdot \vec E = \mu_5 \, \frac{d N_5}{dt} = \frac{e^2\, \mu_5}{2\, \pi^2} \, \int d^3x \, \vec E \cdot \vec {\mathcal B}~.
\end{align}
Taking into account the limiting case of a vanishing electric field $\vec E \to 0 $ (so that the magnitude of $E$ does not depend on $x$, or alternatively assuming a uniform finite electric field) and taking the direction of $\vec E \to 0 $ to be parallel to $\vec {\mathcal B}$, one arrives at the CME, namely the space integral of an electrical current $\int d^3 x \, \vec j_E = \vec J_E$ is proportional to the space integral of the magnetic field $\vec B$ with the proportionality coefficient being given by the chiral chemical
potential $\mu_5$:
\begin{align}\label{chiral3}
\vec J_E = \mu_5 \, \frac{d N_5}{dt} = \frac{e^2\, \mu_5}{2\, \pi^2} \, \int d^3x \, \vec {\mathcal B} = \frac{2\, \alpha}{\pi} \, \mu_5\, \int d^3x \, \vec {\mathcal B}~.
\end{align}
The appearance of the fine structure constant $\alpha = e^2/4\pi$ as a proportionality factor indicates the quantum nature of the phenomenon and is consistent with its connection to the chiral anomaly. In addition, there is an induced chiral current, which is proportional to the chemical potential $\mu$~\cite{cme,mz}:
\begin{align}\label{cme5}
{\vec J}^{\,5} = \frac{e\, \mu}{2\, \pi^2} \, \int d^3x \, \vec {\mathcal B}~.
\end{align}
These effects has been defined in several independent ways in \cite{cme}, including finite temperature $T \ne 0$ formulations, and thus it was argued that CME is independent of temperature for $T$-independent $\mu_5$ and $\mu$~\footnote{It was also argued in \cite{cme} that CME is also independent of the fermion mass, and hence the effect can also characterise massive fermions; however this latter statement may not be correct. There are subtleties 
if a finite fermion mass is present~\cite{dvorn2,dvorn}. At any rate, for our interests here, 
the CME will be studied for high temperatures, above the electroweak phase transition where the fermions are massless.}. 
Such an effect might have important phenomenological implications for the QG plasma physics~\cite{cme}. 

We now come to our case. First we remark that, in string effective actions~\cite{string,kaloper}, the (totally antisymmetric) torsion interpretation of the $H$-field is valid to order $\alpha^\prime$ (quartic in derivatives) 
in an expansion in powers of the Regge slope. In such a torsionful curved space-time, as relevant for our cosmological case, any chiral anomaly that might characterise our system will also involve the generalised (torsionful) Riemann curvature tensor ${\overline R}_{\mu\nu\rho\sigma}(\omega)$ and its dual~\cite{kaloper},
\begin{align}\label{anomalygrav}
\nabla_\mu J^{5\, \mu} = \frac{e^2}{8\, \pi^2} \, F_{\mu\nu} \, \widetilde F^{\mu\nu} - \frac{1}{192\, \pi^2} {\overline R}_{\mu\nu\alpha\beta}(\overline \omega)\, {\widetilde {\overline R}}^{\mu\nu\alpha\beta} (\overline \omega) \equiv {\mathcal G}(A, {\overline \omega})~,
\end{align}
where the overline over a quantity denotes the presence of torsion, and $\overline \omega = \omega + H$ denotes (schematically) the torsionful connection, with $\omega$ the torsion-free connection and $H$ the KR field strength which plays the r\^ole of (totally antisymmetric) torsion. The quantity 
$\nabla_\mu$ denotes the gravitational covariant derivative, with respect to the torsion-free connection. (The reader should notice that there is no $H$-torsion contribution to the covariant four-divergence of a four-vector, due to symmetry reasons.)
The duals are now defined as $\widetilde F_{\mu\nu} = \frac{1}{2} \, \sqrt{-g}\, \epsilon_{\mu\nu\rho\sigma} \, F^{\rho\sigma}$,  and 
${\widetilde {\overline R}}_{\alpha\beta\mu\nu} = \frac{1}{2} \, \sqrt{-g} \, \epsilon_{\mu\nu\rho\sigma} \, {\overline R}_{\alpha\beta}^{\,\,\,\,\,\,\,\,\rho\sigma}$,  
with $g$ the determinant of the metric corresponding to a (torsion-free) Riemann curvature tensor $R_{\mu\nu\rho\sigma}$.
In fact, in differential form notation (which we use below for the sake of brevity), the gravitational part of the anomaly is given by~\cite{kaloper} 
\begin{align}\label{rrtorsion} {\rm Tr}\, \Big(\mathbf{\overline R} (\overline \omega) \wedge \mathbf{\overline R} (\overline \omega) \Big)= 
{\rm Tr}\, \Big(\mathbf{R }(\omega) \wedge \mathbf{R }(\omega)\Big) + {\rm \mathbf{d}}\, \Big[{\rm Tr} \Big(\mathbf{H} \wedge \mathbf{R}) + \mathbf{H} \wedge \mathbf{D} \, \mathbf{H} + \frac{2}{3} \, \mathbf{H} \wedge \mathbf{H} \wedge \mathbf{H} \Big)\Big], 
\end{align}
where $\mathbf{d}$ is the exterior derivative, $\mathbf{D}$ denotes the (torsion-free) gravitational covariant exterior form, 
$\mathbf{D}\, \mathbf{V}^a = \mathbf{d} \mathbf{V}^a + {\omega}^a_{\,\,b}\, \wedge \, \mathbf{V}^b$, 
and the trace  Tr is taken over tangent space (Latin) indices $a,b, ...$, i.e. ${\rm Tr} \, \mathbf{{\overline R}}^a_{\,\,b}(\overline \omega) \wedge \mathbf{{\overline R}}^{b}_{\,\, a}(\overline \omega)$, with $\mathbf{{\overline R}}^{a}_{\,\,b} (\overline \omega) = \frac{1}{2} {\overline R}_{\mu\nu\,\,\,\,b}^{\,\,\,\,\,\,a}\, dx^\mu \wedge dx^\nu = \mathbf{d}\, {\overline \omega}^a_{\, b} + {\overline \omega}^a_c\, \wedge {\overline \omega}^c_{\, b}$,  etc.,  in a standard differential form notation, where the indices $a,b,c \dots$ are raised and lowered by the Minkowski metric.
On the other hand, in order to maintain the conventional $U(1)$ gauge invariance, the Maxwell field strength should be defined as in standard electrodynamics~\cite{kaloper}, with respect to the normal derivative (in form language 
$\mathbf{F}= \mathbf{d A}$, obeying the Bianchi identity 
$\mathbf{d F} =0$).

It is well known~\cite{hull,mavindex} that the torsion contributions (\ref{rrtorsion}) to the anomaly (\ref{anomalygrav}) 
may be removed by the addition of local counterterms (in the standard renormalisation-group sense) to the effective action, provided that the chiral current couples to a gauge field, which is the case of interest here~\footnote{An alternative way to see this, is to observe that the torsion contributions  to the gravitational part of the anomaly, (\ref{rrtorsion}), assume a total exterior derivative (closed) form. This implies that one can appropriately redefine the axial current by such torsion dependent terms~\cite{dobado}, to arrive at a new gauge and Lorentz-invariant current, whose anomaly equation is torsion free.}. 

Hence, the anomaly becomes dependent only on the torsion-free spin connection $\omega$, 
\begin{align}\label{anomalygrav2}
\nabla_\mu J^{5\, \mu} = \frac{e^2}{8\, \pi^2} \, F_{\mu\nu} \, \widetilde F^{\mu\nu} - \frac{1}{192\, \pi^2} {R}_{\mu\nu\alpha\beta}(\omega)\, {\widetilde {R}}^{\mu\nu\alpha\beta} (\omega) \equiv {\mathcal G}(A, {\omega}),~.
\end{align}

This will define a specific form of the low-energy effective action, which we restrict our attention to in this work. 
In fact, as shown in \cite{kaloper}, and discussed briefly in section \ref{sec:intro}, imposing to all orders the constraint on the conservation of the torsion charge (\ref{delta}),  is equivalent to the addition of specific counterterms, which leads to the dual effective action (\ref{seffpi}) in terms of the Lagrange multiplier KR-axion field $b(x)$~\footnote{Some important comments are in order at this point, for clarity and completeness. 
First of all, we stress that in this work we are concerned  with a specific 
kind of (totally antisymmetric) torsion induced by the KR field, $\mathbf{H} = \mathbf{d B}$,  for which one has the Bianchi identity  
$ \mathbf{d} \star \mathbf{H} = 0$, imposed exactly at a quantum level via the constraint (\ref{delta}) (conservation of the torsion charge)~\cite{kaloper}.
It is for this KR torsion that the anomaly has the form (\ref{anomalygrav}). For a generic torsion, however, defined  (in the language of differential forms) as~\cite{torsion}: $\mathbf{T}^a= \mathbf{d} \mathbf{e}^a + \overline \omega^a_b \wedge \mathbf{e}^b = \mathbf{K}^a_b \wedge \mathbf{e}^b$, with $\mathbf{K}$ the contorsion tensor, 
there are~\cite{zanelli} additional topological contributions to the  index of the Dirac operator, and hence the chiral anomaly, which should be added on the right-hand-side of (\ref{anomalygrav}).These extra terms are proportional to the Nieh-Yan topological density~\cite{nie}:
 $ {\mathcal N} = \mathbf{T}^a \wedge \mathbf{T}_a 
-  \mathbf{\overline R} (\overline \omega)_{ab}\, \wedge  \mathbf{e}^a \wedge \mathbf{e}^b  = \mathbf{d} (\mathbf{e}^a \wedge \mathbf{T}_a). 
$ This is a divergent term, requiring proper regularisation, and in \cite{zanelli} it was claimed that any regulator dependence can be properly absorbed in rescalings of the vielbein, so that there are non-trivial contributions of torsion to the anomaly, coming from ${\mathcal N}$. However, there is currently a 
debate~\cite{kreimer} as to whether such contributions can survive the removal of the regulator. (It is worth remarking though that, as a result of the total exterior derivative form of ${\mathcal N}$, its contribution to the anomaly shares a similar fate with that of the H-torsion (\ref{rrtorsion}.) Namely, it can be absorbed  in a redefined current, depending on torsion, whose anomaly equation is torsion free~\cite{dobado}). Fortunately, for our string-inspired KR torsion model such ambiguities do not arise. The Nieh-Yan invariant vanishes identically in this case, due to the Bianchi identity constraint (\ref{delta}); indeed, in terms of the torsion $\mathbf{T}^a$, this constraint can be expressed as $ 0 = \mathbf{d} \star \mathbf{H} = \mathbf{d}(\mathbf{e}_a \wedge \mathbf{T}^a)= {\mathcal N}$, where we took into account that $\star \mathbf{H} \propto \mathbf{e}_a \wedge \mathbf{T}^a $. Upon neglecting the 
Nieh-Yan invariant, then, one can unambiguously find appropriate counterterms~\cite{hull}, taking into account (\ref{rrtorsion}), to express~\cite{mavindex} the index of the torsionful Dirac operator, 
and thus the anomaly (\ref{anomalygrav}), in terms of torsion-free quantities (\ref{anomalygrav2}), as discussed above.}.

Thus, a QED  effective action (in the concrete case the fermions are charged under electromagnetism)  in a space-time with a torsionful connection, is equivalent to a QED action (\ref{seffpi}) in a space time without torsion but with a dynamical KR axion field. The latter contains a dimension six four-fermion operator and a dimension five operator that couples the derivative of the $b$-field to the axial fermion current. By partial integration, then, this dimension-five term in the effective Lagrangian yields a coupling of the $b$ field to the anomaly ${\mathcal G}(A, {\omega})$ (\ref{anomalygrav2}) 
without torsion. Since the torsion contributions to the anomaly are thus ambiguous, removable by an appropriate choice of counterterms, one should not expect any contribution of $B_0$ to the CME, which as discussed above (\emph{cf.} (\ref{chiral2}), (\ref{chiral3})), is linked to the chiral anomaly. This is indeed what happens, as follows from the r\^ole of $B_0$ as an axial background, which is known not to contribute to CME~\cite{kaplan,dvorn}, as we now review 
briefly, for completeness.

To this end, we first remark that, for a Robertson-Walker (RW) cosmological backgound, the $R \widetilde R$ term in (\ref{anomalygrav2}) \emph{vanishes} identically~\footnote{In fact, it is only the gravitational-wave type fluctuations that contribute to the (torsion-free) Riemann-curvature-dependent part of the anomaly (\ref{anomalygrav2}), which can then lead to interesting scenarios for leptogenesis, different from our approach here~\cite{stephon}. Moreover, graviton fluctuations in the $R(\omega) \, \widetilde R(\omega) $ gravitational parts of the anomaly (\ref{anomalygrav2}) might play an important r\^ole in radiative Majorana mass generation for the right-handed neutrinos, as explained in \cite{pilaftsis}. This occurs upon coupling the KR axion to ordinary axion fields, through kinetic mixing, with the ordinary axions coupling in turn to right-handed Majorana neutrinos via axion-shift-symmetry breaking Yukawa couplings. In this sense, the anomaly may be important in generating dynamically the mass scale for the right handed neutrinos $m_N$ itself, that plays a pivotal r\^ole in our leptogenesis scenario~\cite{decesare,bms}.}. Hence, for cosmological RW space-times the axial anomaly is determined only by its gauge-field part. One can therefore discuss the CME in our context by a straightforward extension of the flat space-time case. 

It suffices for our purposes, to restrict our attention to a local frame, where the expansion of the Universe can be ignored (this would be the case if one is interested in examining the effects of $B_0$ on CME during the leptogenesis era, or in a QG plasma~\cite{cme,kaplan} or neutron star situation~\cite{dvorn2}). In the absence of an explicit $\mu_5$ term, i.e. setting $\mu_5=0$ as we have done before, 
we observe from the effective action (\ref{seffbb}), that, at least naively, the background field $B_0 = {\dot {\bar b}}$ seems to play a r\^ole analogous to a chiral chemical potential, which however is in general temperature dependent, as we have discussed in this work. 
If one adds a chemical potential $\mu_5$ term, in order to capture local in space-time effects of the QG plasma, as in \cite{cme}, then, as we have discussed in section \ref{sec:bocond} (\emph{cf.} Eq.~(\ref{fermlag})), this will appear in the combination of an {\emph effective} chemical potential 
\begin{align}
\mu_5^{\rm eff} \equiv \mu_5 - B_0(T),
\label{effcp}
\end{align}
and one \emph{naively} would expect a CME (\ref{chiral3}), with $\mu_5$ replaced by $\mu_5^{\rm eff}$ (\ref{effcp}). 

However, as argued in \cite{kaplan,dvorn}, with different methods between these two works, the axial vector potential $B_0$ does \emph{not} contribute to CME, and instead one has (\ref{chiral3}), even if $B_0 \ne 0$ is present. The subtlety lies in the fact that, in the presence of a background field $B_0$, as we have discussed above (see (\ref{dr})), the dispersion relations for the fermions are affected non-trivially by the presence of $B_0$, which differentiate it from a mere chiral chemical potential, and, moreover, there are subtleties in the ordering of taking the massless limit $m \to 0$. 
As discussed in \cite{dvorn}, it is important that we take the massless (chiral) limit $m \to 0$ at the end of the computation, that is first, one should assume massive fermions, in the presence of a $B_0 \ne 0$, solve the corresponding Dirac equation, and only at the end take the limit $m \to 0$. Should had one started, instead, with the chiral Lagrangian for {\emph massless} quarks from the beginning, and then turned on an external vector time-like potential, $B_0 \ne 0$, the appearance of $B_0$ contributions to the CME (through (\ref{effcp}))  would have occurred, which, however, would not be correct  from the point of view of energy conservation~\cite{kaplan}. 

Let us briefly review the situation below, following the relativistic quantum mechanics analysis of \cite{dvorn}, adopted to our case.  Let us consider for concreteness the Dirac equation for a charged fermion of mass $m$, which initially is taken \emph{non zero}, in an external homogeneous and isotropic (\emph{i.e.} constant) magnetic field, of intensity ${\mathcal B}$, along the $z$ direction, and a time-like axial potential $B_0 \ne 0$, assumed also constant for brevity (any mild temperature dependence, as is our cosmological case, will not affect the conclusions). The Dirac Lagrangian is given by (\ref{fermions2}) with the additional term describing the coupling of the charged fermions to the external electromagnetic potential, $A_\mu = (0,\,0,\, {\mathcal B}\, x, \, 0)$, corresponding to the (constant) magnetic field of intensity ${\mathcal B}$ along the $z$ direction:
\begin{align}\label{chargedfermions}
S_{Dirac} = &\; \int d^{4}x\sqrt{-g}\bar{\psi}\Big(\gamma^{\mu}(\imath\, \partial_{\mu}  + q_e \, A_\mu) - m\Big)\psi - \int d^{4}x\sqrt{-g}\, B_{0} \, \bar{\psi}\, \gamma^{0} \,  \gamma^{5}\psi \,.
\end{align}

The detailed solution of the Dirac equation of \cite{dvorn,rqm}, yields for the wave function of the fermion
\begin{align}\label{psisol}
\psi (t,x,y,z) = e^{-iE t + i p_y y + i p_z z}\, {\widetilde \psi}(x),  
\end{align}
where the bispinor $\tilde \psi (x)$ is given by~\cite{dvorn}
\begin{align}\label{psitilde}
\widetilde \psi (x) =\begin{pmatrix} &C_1 u_{n-1}(\xi)  \\ & \imath C_2 u_n (\xi) \\ &C_3 u_{n-1} (\xi)\\ & \imath C_4  u_n (\xi)\end{pmatrix}, \quad u_n (\xi) = \Big(\frac{q_e\, {\mathcal B}}{\pi}\Big)^{1/4} \, e^{-\frac{\xi^2}{2}} \, \frac{H_n(\xi)}{\sqrt{2^n\, n!}}, \,\, n=0,1, \dots, \quad \xi= \sqrt{q_e \, {\mathcal B}} \, x + \frac{p_y}{\sqrt{q_e \, {\mathcal B}}},
\end{align}
with $H_n (\xi)$ Hermite polynomials. The constants (``spin coefficients''~\cite{dvorn,rqm}) $C_i, \, i=1,\dots 4,$ obey the equations (for concreteness, below we restrict ourselves to particles):
\begin{align}\label{system}
(-{E} \mp p_z \mp B_0)C_{1,3}  \mp \sqrt{2\,q_e\, {\mathcal B} \, n} \, C_{2,4} + m \, C_{3,1} &= 0, \nonumber  \\
(-{E} \pm p_z \mp B_0)C_{2,4}  \mp \sqrt{2\,q_e\, {\mathcal B} \, n} \, C_{1,3} + m \, C_{4,2} &= 0. 
\end{align}
As follows from (\ref{system}), the energy levels are split in Landau levels, parameterised by a non-negative  integer $n=0,1,2 \dots ,$ and in the presence of $B_0 > 0$ (in our case and in our conventions) they are obtained from the dispersion relation~\cite{dvorn,rqm}:
\begin{align}\label{energies}
E^2 =  ({\mathcal E}_0 - s B_0)^2 + m^2~,   \quad  {\mathcal E}_0  = \sqrt{p_z^2 + 2\, q_e\, {\mathcal B} \, n},     \quad n = 0, 1, \dots , \quad s=\pm1,  \nonumber \\
\end{align}
with $p_z$ the longitudinal momentum of the (charged) fermion along the magnetic field, $q_e$ is the electric charge,  $s =\pm 1$ is the discrete quantum number associated with the spin operator'~ \cite{dvorn}). When taking the square root in (\ref{energies}) to evaluate the energy  $E$, 
the overall sign of $E$ is associated with particles (+) or antiparticles (-). We shall consider for our purposes here the lowest Landau level $n=0$, as the higher levels do not contribute to the induced CME~\cite{dvorn}. In that case the energy (\ref{energies}), can be written as
\begin{align}\label{energiesn0}
E^{(n=0)}=\pm \sqrt{(|p_z| \mp B_0)^2 + m^2} = \pm \sqrt{(p_z - B_0)^2 + m^2}
\end{align}
where, in the last equality on the right-hand side, the two values of $s=\pm1$ have been absorbed in $|p_z|$, which now can be written as (the algebraically valued) $p_z$. 
The reader should notice that, in the case of the lowest level $n=0$, one has \emph{one spin state} of the charged fermion~\cite{dvorn,rqm}. In the massless limit $m \to 0$, one has from (\ref{energiesn0}), for $n=0$, 
\begin{align}\label{masslessdr}
E^{(n=0)} = \pm | p_z - B_0 | \equiv \pm {\mathcal E},
\end{align}
which the reader should compare with the $m \to 0$ limit of the dispersion relation (\ref{dr}), where the r\^ole of $s(=+1)$ is played by the  helicity $\lambda (=+1)$, the latter defined as the projection of the spin vector in the direction of momentum. Here the direction of the momentum is parallel to the magnetic field, which has been taken to lie along the $z$-axis.

Considering the $n=0$  and $m\to 0$ limiting case in (\ref{system}), we then observe that, in order to avoid solutions with negative-index Hermite polynomials, 
one should set $C_1=C_3=0$, which implies:
\begin{align}\label{system2}
(-{\mathcal E} + p_z - B_0)\, C_{2} = 0 \,  \Rightarrow  \, {\mathcal E}= p_z - B_0, \quad C_2 \ne 0, \, C_4=0,
\end{align}
\begin{align}\label{system3}
(-{\mathcal E} - p_z + B_0) \, C_{4}  = 0 \, \Rightarrow  \, {\mathcal E}= -p_z + B_0, \quad C_4 \ne 0, \, C_2=0.
\end{align}
For further details into the form of the wave-functions of the solution and the non-trivial spin coefficients we refer the reader to ref.~\cite{dvorn} (and \cite{rqm}).
The solution (\ref{system2}) ((\ref{system3})) can be identified with the left- (right-) handed fermion. Comparing with (\ref{masslessdr}) (for the particle (+) case, to which we restrict ourselves below for concreteness - the corresponding relations for antiparticles are obtained straightforwardly from (\ref{energies}) upon taking the $-$ sign for the overall energy), we then observe, that for a right-handed fermion particle one has 
\begin{align}
|p_z - B_0 | = -p_z +B_0 \, \Rightarrow \, p_z \, < \, B_0,
\end{align}
while for a left-handed fermion, 
\begin{align}
|p_z - B_0 | = p_z - B_0 \, \Rightarrow \, p_z \, > \, B_0\, >0.
\end{align}
These relations imply that the left-handed fermion particle total energy at the lowest Landau level $n=0$, reads:
\begin{align}\label{el}
E_{L}^{(n=0)}= p_z - B_0 , \quad   0 \, < \, B_0 \, < \, p_z \, < \, +\infty,
\end{align}
while for the right-handed-fermion energy one has
\begin{align}\label{er}
E_{R}^{(n=0)}= -p_z + B_0 , \quad   -\infty \, < \, p_z \, < \, B_0.
\end{align}
The range of the momenta $p_z$ is important, and in fact in \cite{dvorn2}, this range was different, with its finite bounds being zero. This lead erroneously to $B_0$ contributions to the CME (\ref{chiral3}). This is not the case with the momentum range (\ref{el}), (\ref{er}). Indeed, let one calculate the induced electrical   current density 
as a thermal equilibrium ensemble~\cite{dvorn} 
\begin{align}\label{enscurr}
\vec j^{\,(n=0)} = \vec j_L^{\,(n=0)} + \vec j_R^{\,(n=0)}, 
\end{align}
with 
\begin{align}
\vec j_{L,R}^{\,(n=0)} = -q_e \, \int dp_y\, dp_z \,  \overline \psi_{L,R}\, \vec \gamma \, \psi_{L,R} \, f(E_{L,R}^{(n=0)} - \mu_{L,R}),
\end{align}
where $f(E) = \Big( e^{E/T} + 1 \Big)^{-1}$ is the Fermi-Dirac distribution at a temperature $T$, and $\mu_{L(R)}$ are the chemical potentials of the left (right) particles, with $\mu_{L(R)} = \mu -(+) \mu_5$ the chemical potentials for left(right)-handed fermions, $\psi_{L(R)}$, with $\mu_5$ the chiral chemical potential, and $\mu$ the chemical potential for the (non-chiral) fermions $\psi$.
The wave-functions $\psi_{L,R}$ are obtained from (\ref{psisol}),(\ref{psitilde}) in the chiral limit $m \to 0$, upon replacing the energy $E$ by $E_{L,R}$, respectively. Details of the solutions are given in \cite{dvorn}, and will be omitted here. 
We only give below the final result for the average particle current (\ref{enscurr}) (it can be shown~\cite{dvorn} that the higher Landau levels $n \ne 0$ contributions to the current are zero, hence from now we omit the superscript $(n=0)$ in the notation for the current, for brevity):
\begin{align}\label{cmeaxial}
\vec j &= \frac{q_e^2\, \vec {\mathcal B}}{4\, \pi^2} \, \Big[ \int_{-\infty}^{B_0} dp_z \, f(-p_z +B_0 - \mu_R) - \int^{+\infty}_{B_0} dp_z \, f(p_z -B_0 - \mu_L) \Big] \nonumber  \\
&= \frac{q_e^2\, \vec {\mathcal B}}{4\, \pi^2} \,  \int^{+\infty}_{0} d\tilde p \, \Big(f(\tilde p- \mu_R) - f(\tilde p - \mu_L) \Big),
\end{align}
where, in order to arrive at the result given in the second line of (\ref{cmeaxial}), we have changed the integration variable $p_z - B_0 \to \tilde p$ (and as far as the first integral of the right-hand-side of the first line is concerned, we also made the change $\tilde p \to -\tilde p$). The right-hand-side of (\ref{cmeaxial}) is independent of $B_0$. 
In a similar way one calculates the antiparticle current $\vec{\overline J}$, taking into account that for antiparticles $ \mu_{L,R} \to -\mu_{R,L}$:
\begin{align}\label{cmeaxial2}
\vec{\overline j}  = \frac{q_e^2\, \vec {\mathcal B}}{4\, \pi^2} \,  \int^{+\infty}_{0} d\tilde p \, \Big( f(\tilde p+ \mu_L) - f(\tilde p+ \mu_R)\Big) ,
\end{align}
which is also independent of $B_0$. The total particle plus antiparticle current is independent of $B_0$, and given by 
\begin{align}\label{cmetotalaxial} 
 \vec j + \vec{\overline j} &= \frac{q_e^2\, \vec {\mathcal B}}{4\, \pi^2} \, \int^{+\infty}_{0} d\tilde p \,  
 \Big(f(\tilde p- \mu_R)   - f(\tilde p+ \mu_R) 
 -  f(\tilde p - \mu_L) +  f(\tilde p+ \mu_L) 
\Big) = \frac{2\alpha}{\pi}\, \mu_5 \, \vec {\mathcal B},
 \end{align}
with $\alpha = q_e^2/4\pi$ (which coincides with the fine structure constant if $q_e=e$, the fundamental electric charge). This
is the standard CME (\ref{chiral3}), being proportional only to the chiral chemical potential, and \emph{independent} of the axial-vector background $B_0$ and temperature. This is in agreement with the  result of \cite{kaplan}, derived using energy conservation arguments, and also with the general expectations from the theory of quantum anomalies, as discussed above.

.

\section*{Acknowledgements}

We thank L.C. Garcia de Andrade for discussions. The work of TB is supported by an STFC (UK) research (doctoral) studentship and that of NEM and SS is supported in part  by STFC (UK) under the research grant ST/P000258/1. N.E.M. also acknowledges a scientific associateship (``\emph{Doctor Vinculado}'') at IFIC-CSIC-Valencia University\ (Spain).

\end{document}